\documentclass[twocolumn,prl,amsmath,amssymb,showpacs,superscriptaddress]{revtex4-1}
\usepackage{epsf}
\usepackage{graphicx}
\usepackage{color}
\usepackage{gensymb}
\usepackage{amsmath}
\usepackage{multirow}
\usepackage{eqnarray}
\usepackage{sidecap}
\usepackage[normalem]{ulem}

\begin{document}

\title {X-ray Absorption Spectroscopy Study of La$_{1-y}$Sr$_{y}$Co$_{1-x}$Nb$_x$O$_3$}
\author{Rishabh Shukla}
\affiliation{Department of Physics, Indian Institute of Technology Delhi, Hauz Khas, New Delhi-110016, India}
\author{Ajay Kumar}
\affiliation{Department of Physics, Indian Institute of Technology Delhi, Hauz Khas, New Delhi-110016, India}
\author{Ravi Kumar}
\affiliation{Atomic and Molecular Physics Division, Bhabha Atomic Research Centre, Mumbai-400085, India}
\author{S. N. Jha}
\affiliation{Beamline Development and Application Section, Physics Group, Bhabha Atomic Research Centre, Mumbai-400085, India}
\author{R. S. Dhaka}
\email{rsdhaka@physics.iitd.ac.in}
\affiliation{Department of Physics, Indian Institute of Technology Delhi, Hauz Khas, New Delhi-110016, India}
\date{\today}

\begin{abstract}

We use x-ray absorption spectroscopy to investigate the local structure and electronic properties of bulk La$_{1-y}$Sr$_{y}$Co$_{1-x}$Nb$_x$O$_3$ ($y=$ 2$x$ as LSCNO and $y=$ 0 as LCNO) samples. The x-ray absorption near-edge spectra (XANES) of LSCNO at Co K-edge affirm the valence state of Co in 3+. However, in the case of the LCNO, a subtle variation in the valence state of Co ions from 3+ to 2+ is evident with Nb substitution. The detailed analysis of the Fourier transform (FT) of the extended x-ray absorption fine structure (EXAFS) for the LSCNO samples exhibit the two groups of the bond-lengths owing to the Jahn-Teller (JT) distortion in the CoO$_6$ octahedra, which manifest that the Co$^{3+}$ ions exist in the intermediate spin-state (t$^5_{2g}$e$^1_g$) at room temperature. However, we find that the JT distortion is not present in LCNO samples for $x>$0.025 due to an increase in the high-spin Co$^{2+}$/Co$^{3+}$ ions accompanied by the Nb--induced structural transformation. Intriguingly, the La L$_3$-edge spectra for the LSCNO samples exhibit that the La ions exist in the trivalent state and the local disorder around La atoms decreases with Sr and Nb substitution. Interestingly, the simulated FT of the EXAFS spectra at the La L$_3$-edge demonstrates the three groups of La--O bond lengths, which exhibit a monotonous change with Sr and Nb substitution. Moreover, the XANES measured at Sr and Nb K-edges confirm their oxidation state to be in 2+ and 5+, respectively.
\end{abstract}

\maketitle

\section{\noindent ~Introduction}

The intriguing physical properties of complex perovskite oxides have been investigated over a few decades for their advance technological as well as environmental applications \cite{BednorzRMP88, GrangerWilley15, KubicekJMCA17, SimbockNC20}. The family of perovskite oxides has a general chemical formula of ABO$_3$, where A is a bi/trivalent rare-earth/alkali-earth cation and B is mostly a transition metal cation. Here, the physical properties are mostly governed by the BO$_6$ octahedral coordination, where the ionic radii and valence state of constituent entities play a key role \cite{GoodenoughJSSC58}. The close-packed structure of these perovskites led to defining a tolerance factor as $\tau = ({r_{\rm A}+r_{\rm O}})/{\sqrt{2}(r_{\rm B}+r_{\rm O})}$ named after Goldschmidt, where $r$ denotes the ionic radii of the subscript ion \cite{GrangerWilley15}. The specific values of $\tau$ are determinants of different families of crystal structures and accompanied with octahedral rotation as defined by Glazer \cite{GlazerAC72}, which is crucial in understanding the correlation between the physical properties of materials and their structural fingerprints. In the family of complex oxides, the RCoO$_3$ (R = rare-earth elements) cobaltites amidst perovskites comprise a large family, which exhibit interesting phenomena predominantly originating from the interplay of different spin-states of Co ion \cite{IvanovaPU09}, which can exist in 2+, 3+, and/or 4+ valence states. More specifically, in the case of the LaCoO$_3$ (LCO) sample, the spin-state transition, anomalous magnetism and transport properties have been explored extensively using theoretical as well as experimental tools \cite{PandeyPRB08} to investigate the effect of different external perturbations, like temperature, pressure, magnetic field, strain, cationic substitution, etc. \cite{KorotinPRB96, VankoPRB06, HsuPRB09, HaverkortPRL06, ShuklaPRB18, BaierPRB05, VoigtPRB03, ShuklaTSF20, AbbatePRB93, KumarPRB101}. It has been well established that the LCO exhibits a diamagnetic ground state owing to the complete pairing of electrons in the t$_{2g}$ states of Co$^{3+}$(3d$^6$) ions and termed as low spin (LS, t$^6_{2g}$e$^0_g$, S=0) state \cite{RaccahJAP68, LamPRB80}. This LS state evolves with the temperature due to the small difference between crystal field splitting ($\Delta_{cf}$) and Hund's exchange interaction energy (J$_{ex}$). A partial transition into an intermediate spin (IS, t$^5_{2g}$e$^1_g$, S=1) state and/or high spin (HS, t$^4_{2g}$e$^2_g$, S=2) state appears near 85~K \cite{KreinerPRB04, SaitohPRB97, RadaelliPRB02}, where strong competition between the $\Delta_{cf}$ and J$_{ex}$ is considered to be responsible for this spin-state transition \cite{BarmanPRB94, YangPRB15, NekrasovPRB03}. Further, an evolution of these states is evident near 500~K, where an insulator to metal transition is perceived in different measurements {\cite{BaierPRB05, RadaelliPRB02, ZobelPRB02}. Among the various spin-states of Co in the LCO, only IS state is expected to be strongly Jahn-Teller (JT) active, as predicted in the references \cite{KorotinPRB96, LamPRB80, ZobelPRB02}. The presence of JT distortion in the CoO$_6$ octahedra lowers the crystal symmetry and result in the two different groups of Co--O bond-lengths \cite{KorotinPRB96, BaierPRB05, YamaguchiPRB96}. In this context, x-ray absorption spectroscopy (XAS) has emerged as an advanced tool to probe the change in the local structure and electronic properties \cite{WuPRB97, GrootCR01}. The overlap between atomic orbitals of the absorbing atom and the neighboring atoms are sensitive to the bond-lengths and attribute to the near and far edge features of absorption spectra. Interestingly, the cationic substitution in perovskites alters the bond-lengths and bond-angles, which lead to the change in the atomic orbital overlap and hence tailor their electronic structure. Moreover, the JT effect has also been investigated in Co/Mn-based perovskite oxides using XAS measurements \cite{JiangPRB07, DownwardPRL05, JiangPRB09, LoucaPRL03, ThorntonJPCM91, HaasJSSC04, ToulemondeJSSC01, SundaramPRL09}.

Recently, we have reported the structural transformation in LaCo$_{1-x}$Nb$_x$O$_3$ from rhombohedral to orthorhombic at $x=$ 0.10 and then to monoclinic for $x=$ 0.15, where the substitution of Nb$^{5+}$ ion at Co$^{3+}$ site gradually alters the valence state of Co ions from 3+ into 2+ \cite{ShuklaPRB18}. The change in the crystal structure is accompanied by the genesis of unconventional magnetic behavior, where spin-orbit interaction dominates due to the increased concentration of Co$^{2+}$ ions having a $^4F$ ($^4T_1$) ground state \cite{LloretICA08, KumarPRB102}. On the other hand, the Sr substitution at La site in LCO up to $x=$ 0.50 concentration does not alter the crystal structure, but a ferromagnetic metallic ground state emerges for $x\ge$ 0.18 \cite{SenarisJSSC95, RaviJALCOM18}. These fascinating results obtained with the Sr and Nb substitution derived our interest in the co-substitution of Sr and Nb ions simultaneously at La and Co sites in LCO with a stoichiometric ratio of 2:1, respectively, to preserve the valence state of Co ions in 3+ only \cite{ShuklaJPCC19}. Therefore, to get a detailed insight into the interesting physical properties of La$_{1-y}$Sr$_{y}$Co$_{1-x}$Nb$_x$O$_3$, it is vital to explore the local structure and electronic properties using x-ray absorption spectroscopy. The element-specific x-ray absorption spectra measured with high energy and fine resolution at synchrotron sources led to extract the precise quantitative information about the correlation between electronic and geometric structures \cite{GrootCR01}. The K-edge absorption spectra of any element consist of four integral components: (i) the most prominent dipolar transition ($\Delta l=\pm1$) from the core 1$s$ to the unoccupied $p$ states is termed as absorption edge, which is related to the valence state of the absorbing element as its peak position depends on the binding energy of the atom, (ii) the other part at a lower energy of the edge consists of the local electronic information, i.e., this pre-edge region gives the information about the symmetry of absorbing atom and local electronic structure (these features contain lower intensity owing to the contributions from the quadrupole ($\Delta l=\pm$ 2) transitions), (iii) the region within 50~eV of the main absorption edge is x-ray absorption near-edge spectra (XANES), which is useful to get information about the electronic structure, and (iv) the analysis of higher energy extended x-ray absorption fine structure (EXAFS) region after the 50~eV gives information about the geometric structure and neighboring atoms' position. 

Therefore, herein, we report an investigation of the local structure and electronic properties in La$_{1-y}$Sr$_{y}$Co$_{1-x}$Nb$_x$O$_3$ ($y=$ 2$x$ as LSCNO and $y=$ 0 as LCNO) using the x-ray absorption spectroscopy. We have recorded the absorption spectra at Co K-, La L-, Sr and Nb K-edges at room temperature and analyzed the data in the XANES and/or EXAFS regions. The detailed analysis of the Fourier transform of the EXAFS spectra at the Co K-edge for the LSCNO samples exhibit that Co$^{3+}$ ions in the IS state are JT active with two sets of Co--O bond-lengths, and $x=$ 0.025 sample in LCNO manifests the JT effect in contrast to its magnetic results. The La L$_3$-edge absorption spectra confirm that La ions preserve their trivalent state in all the samples, and a decrease in the value of full width at half maximum for the white line of absorption spectra reveal the decrease in the local disorder with cationic substitution. The EXAFS analysis for La L$_3$-edge exhibit three sets of La--O bond lengths and measurements at the Sr and Nb K-edge confirm their oxidation state to be in the 2+ and 5+, respectively.

\section{\noindent ~Experimental Section}

Polycrystalline samples of La$_{1-y}$Sr$_{y}$Co$_{1-x}$Nb$_x$O$_3$ ($y =$ 2$x$ denoted as LSCNO and $y =$ 0 as LCNO, where $x=$ 0--0.15) were synthesized by the conventional solid-state method, more details of preparation and characterization can be found in references \cite{ShuklaPRB18, ShuklaJPCC19}. We use energy scanning EXAFS beamline (BL-09), which operates in a large energy range of 4--25~keV at Indus--2 synchrotron source (2.5 GeV, 300 mA) at Raja Ramanna Centre for Advanced Technology (RRCAT), Indore, India. We have performed hard x-ray absorption spectroscopy measurements at room temperature to record element-specific absorption edges (Co K, La L, Sr, and Nb K) in the transmission mode. In the transmission mode, precise measurements of the absorption spectra are facilitated by the three different ionization chambers (300~mm length each). The first, second, and third ionization chambers measure the incident flux (I$_0$), transmitted flux (I$_t$), and the spectra of reference metal foils to perform the energy calibration of the monochromator, respectively. The absorption of the x-ray beam is obtained by measuring the flux with and without the sample as denoted by the I$_0$ and I$_t$, respectively, using the relation $I_t$ = $I_0$ exp$^{-\mu t}$, where $\mu$ is the absorption coefficient and $t$ is the absorber thickness. Further details of this beamline can be seen in the reference \cite{BasuJPCS14}. 

A detailed analysis of x-ray absorption spectra has been carried out using the IFEFFIT software \cite{RavelJSR05}. The EXAFS curve fitting analysis is performed in the R-space to the real and imaginary parts of the Fourier transform of $\chi(k)$, using the EXAFS equation given as \cite{KoningsbergerJWS88, SternNH83, RehrRMP00},
\begin{eqnarray*}
 \chi(k) &=& \sum_{shell~j}  -S_0^2\frac{N_j|f_j(\pi, k)|}{kR_j^2}\\
&&\times sin[2kR_j+2\delta_j(k)]e^{-\frac{2R_j}{\lambda(k)}} e^{-2\sigma_j^2k^2}
\end{eqnarray*}
where S$_0^2$ is the many-body amplitude reduction factor ($0\le S_0^2 \le1$), R$_j$ is the interatomic distances (distance between the absorber and scatterer), N$_j$ is the number of scatterers in the j$^{th}$ shell, $\sigma_j$ is the temperature-dependent RMS fluctuations in the bond-length called as $\textquoteleft$Debye-Waller' (DW) factor, $f_j$ is the backscattering amplitude of each neighbor N$_j$ of type $j$ atom, $\delta_j$ is the central-atom partial wave phase-shift of the final state and $\lambda(k)$ is the energy-dependent mean-free path. The above equation can give convenient information from the EXAFS curve fitting analysis; however, the k-space signal has different constituent frequencies and therefore the Fourier transform of the data is required for further visualization and in-depth analysis. The EXAFS spectrum contains information for more than one frequency and depends on the complex variation in amplitude. Therefore, the Fourier transform of an EXAFS spectrum results in the pseudo radial distribution function. The Fourier transform of the k-space signal is in terms of radial distance R and represented by the real function $\chi(R)$. The peaks in $\chi(R)$ can be related to the bond-lengths from neighboring atoms and the equation can be written as \cite{KoningsbergerJWS88, SternNH83, RehrRMP00}, 
\begin{eqnarray*}
\chi(R) = \frac{1}{\sqrt{2\pi}} \int_{k_{min}}^{k_{max}} \omega(k) k^n \chi(k) e^{-2kR} dk\\
\end{eqnarray*}
where $\chi(R)$ is the Hanning window function integrated between the defined k window (k$_{min}$ and k$_{max}$ are the minimum and maximum values of transformed k-space), $\omega(k)$ is the Gaussian window function, and k$^n$ is the weight factor ($n=$ 0, 1, 2, 3). 

\section{\noindent ~Results and Discussion}

\subsection{\noindent ~XANES Analysis}

\begin{figure}
\centering
\includegraphics[width=3.5in]{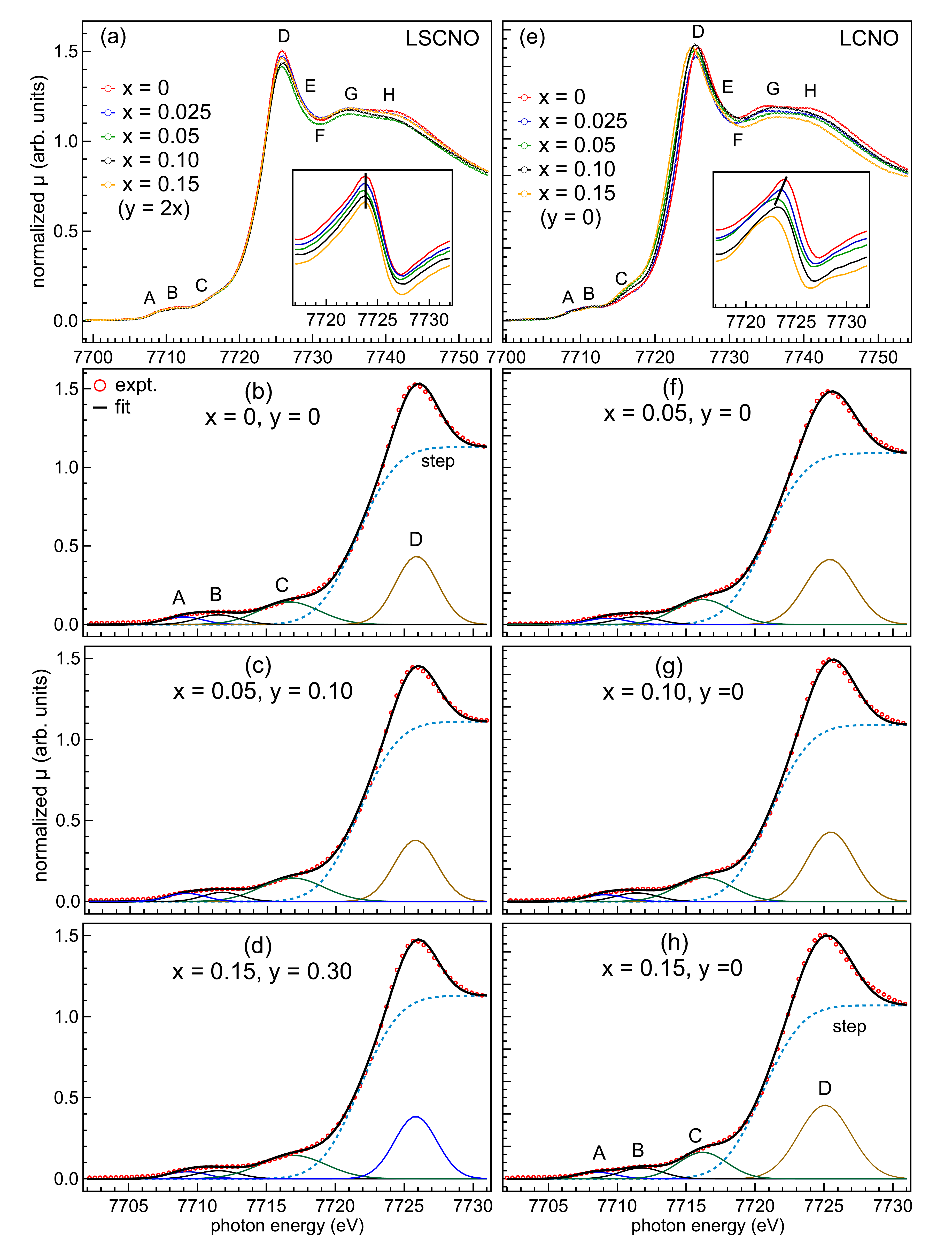}
\renewcommand{\figurename}{Figure}
\caption{The normalized Co K-edge XANES spectra of the bulk La$_{1-y}$Sr$_{y}$Co$_{1-x}$Nb$_x$O$_3$ samples at room temperature (a) LSCNO and (e) LCNO ($x=$ 0--0.15), where different features are marked with eight alphabets (A--H). The insets in (a, e) show the variation in the first-order differential around the edge jump to highlight the change in the valence state of Co ions. The deconvoluted and fitted XANES spectra with four components and one step function for the (b--d) LSCNO and (f--h) LCNO samples.}
\label{fig:XANES_Co_K}
\end{figure}

The room temperature normalized XANES spectra measured at Co K-edge are presented in Figs.~\ref{fig:XANES_Co_K}(a, e) for the LSCNO and LCNO samples, respectively. The energy scale has been calibrated with the cobalt metal-foil by adjusting the Co K-edge at 7709~eV and utilizing a maximum in the first-order derivative of $\mu$. For further analysis, we perform a linear baseline correction in the pre-edge range and a spline baseline correction to the post-edge range in the Athena software \cite{RehrRMP00, RavelJSR05}. These baseline-corrected data were normalized to unity by using a point 800~eV away from the absorption K-edge of Co, similar to the ref.~\cite{HaasJSSC04}. The XANES of these samples exhibit eight distinct observable features highlighted by the alphabets A--H in Figs.~\ref{fig:XANES_Co_K}(a, e). The origin of first two features A and B have discrepancies in the literature and several reports have accredited them to the quadrupole transition between Co 1$s$$\rightarrow$3$d$, which were ascribed as the presence of e$_{\rm g}$ and t$_{\rm 2g}$ states, respectively \cite{HaasJSSC04, ToulemondeJSSC01, MedardePRB06}. However, in a centrosymmetric environment, the 1$s$$\rightarrow$e$_{\rm g}$ and 1$s$$\rightarrow$t$_{\rm 2g}$ transitions are not allowed in the dipole approximation ($\Delta l=\pm$1), though they are allowed in the quadrupolar transitions because of the hybridization between 4$p$ and 3$d$ orbitals  \cite{MedardePRB06, WestreJACS97, YamamatoXRS08}. Interestingly, the recent studies following the charge-transfer multiplet calculations manifest that the first feature A is related to the quadrupolar transition from the 1$s$ to 3$d$ states, while the second feature B is accredited to the dipolar transition from 1$s$ to the intersite hybridization of O 2$p$ mediated Co 3$d$ and Co 4$p$ states \cite{VankoArxiv08}. This explanation of the pre-edge feature is further validated by the Sterbinsky {\it et al.} in the angular XANES spectra of LCO epitaxial thin films \cite{SterbinskyPRB12}.} 

Further, feature C is related to the transition from the Co 1$s$ to Co 4$p$ and La 6$p$ hybridized orbitals and can be assigned to the main transition (1$s$$\rightarrow$4$p$) followed by the ligand to metal charge-transfer (LMCT) shakedown process (1$s^1\bar{c}$3$d^6L^2$ $\rightarrow$ 1$s^1\bar{c}$3$d^7\bar{L}^1$), where $\bar{c}$ is core-hole and $\bar{L}$ is the ligand hole as explained in the earlier reports \cite{PandeyJPCM06, ChainaniPRB92, KimPB97}. The highest intensity main edge feature D is assigned to the 1$s$$\rightarrow$4$p$ dipole transition \cite{ChangPB03}. Moreover, the features E and F manifest the transitions from the Co 4$p$ states hybridized with La 6$p$ and O 2$p$ orbitals; however, for feature F contribution of the La 6$p$ state is higher as compared to the O 2$p$ states \cite{PandeyJPCM06}. The additional features G is related to the hybridization of La 6$p$, Co 4$p$ with O 2$p$ states and H is accredited with the Co 4$p$ state hybridized with O 2$p$ \cite{PandeyJPCM06}. The theoretical calculations by Pandey {\it et al.} predicted that the e$_g$ orbitals are strongly hybridized with O 2$p$ orbitals and hence their atomic character is affected. The higher admixture with O 2$p$ orbitals lowers the Co e$_{\rm g}$ atomic character and will diminish the intensity of the pre-edge feature \cite{PandeyJPCM06}, which is directly proportional to the Co--O bond lengths in these samples.

Interestingly, in the case of the LSCNO samples, we observe no significant shift in the main peak position (feature D, 7726~eV) of the normalized XANES, as clearly visible in the first derivative of edge jump, shown in the inset of Fig.~\ref{fig:XANES_Co_K}(a). By comparing the peak position with the standard reference spectra of Co$_2$O$_3$ \cite{SikoraPRB06} as well as with other reports on LCO \cite{PandeyJPCM06, ChangPB03, HaasJSSC04}, we found that the Co ions preserve their oxidation state of 3+ in all LSCNO samples. This indicate the invariance in the oxidation state of Co with the co-substitution of Sr and Nb in 2:1 ratio, which is also consistent with our recent study in ref.~\cite{ShuklaJPCC19}. Note that all the LSCNO samples are present in the rhombohedral structure (space group, R-3c) up to the $x=$ 0.15 \cite{ShuklaJPCC19}. On the other hand, XANES spectra of LCNO samples are presented in Fig.~\ref{fig:XANES_Co_K}(e), which exhibit a shift in the main peak position (feature D, 7726~eV) and the rising edge towards the lower photon energy, as also articulately seen in the first derivative of edge jump in the inset of Fig.~\ref{fig:XANES_Co_K}(e). This shift in the rising edge of XANES towards the lower photon energy is due to the shielding effect with a decrease in the average valence state of Co ions, i.e., an increase in the overall negative charge of the atom \cite{SinghJAP14, CuarteroPRB16}, which affirm the appearance of Co$^{2+}$ ions due to Nb$^{5+}$ substitution at the Co$^{3+}$ site in LCO \cite{ShuklaPRB18}. Notably, a shift of about 3~eV is reported for the change in the oxidation state of Co from 3+ to 2+ in La$_{1-x}$Mn$_x$CoO$_3$ and other similar materials \cite{SikoraPRB06, HaanJMMM08, IgnatovPRB01}. Interestingly, the LCNO samples possess a rhombohedral structure up to $x=$ 0.05 and a shift of $\approx$0.85~eV towards the lower energy (highlighted by a solid black line in the first derivative of edge jump, see the inset of Fig.~\ref{fig:XANES_Co_K}(e)), which manifest that Nb substitution alters Co oxidation state gradually from 3+ into 2+. For the $x=$ 0.05 sample, the calculated theoretical average valence state of Co ion is 2.75 (the balanced charge equation for Nb substitution in LCO can be written as LaCo$^{3+}_{1-3x}$Co$^{2+}_{2x}$Nb$^{5+}_{x}$O$_3$ \cite{ShuklaPRB18}), which advocates a shift of $\approx$0.75~eV. Here, a small difference between the experimental and calculated values can be related to the fact that the shift of an edge in the oxides is governed by (i) average bond-length (Co--O bond), and (ii) transferred charge \cite{BenfattoPRL99, ElfimovPRL99}. These two effects are correlated to each other such that the decrease in valence state is accompanied by the increase in the bond-lengths. Moreover, the LCNO samples with $x >$ 0.05 exhibit structural transformation and will have an increased concentration of Co$^{2+}$. Therefore, a combined effect of structural transition and oxidation state is observed in the LCNO samples \cite{WaychunasAM87, KuoJPCB08}. 

\begin{figure}
\centering
\includegraphics[width=3.5in]{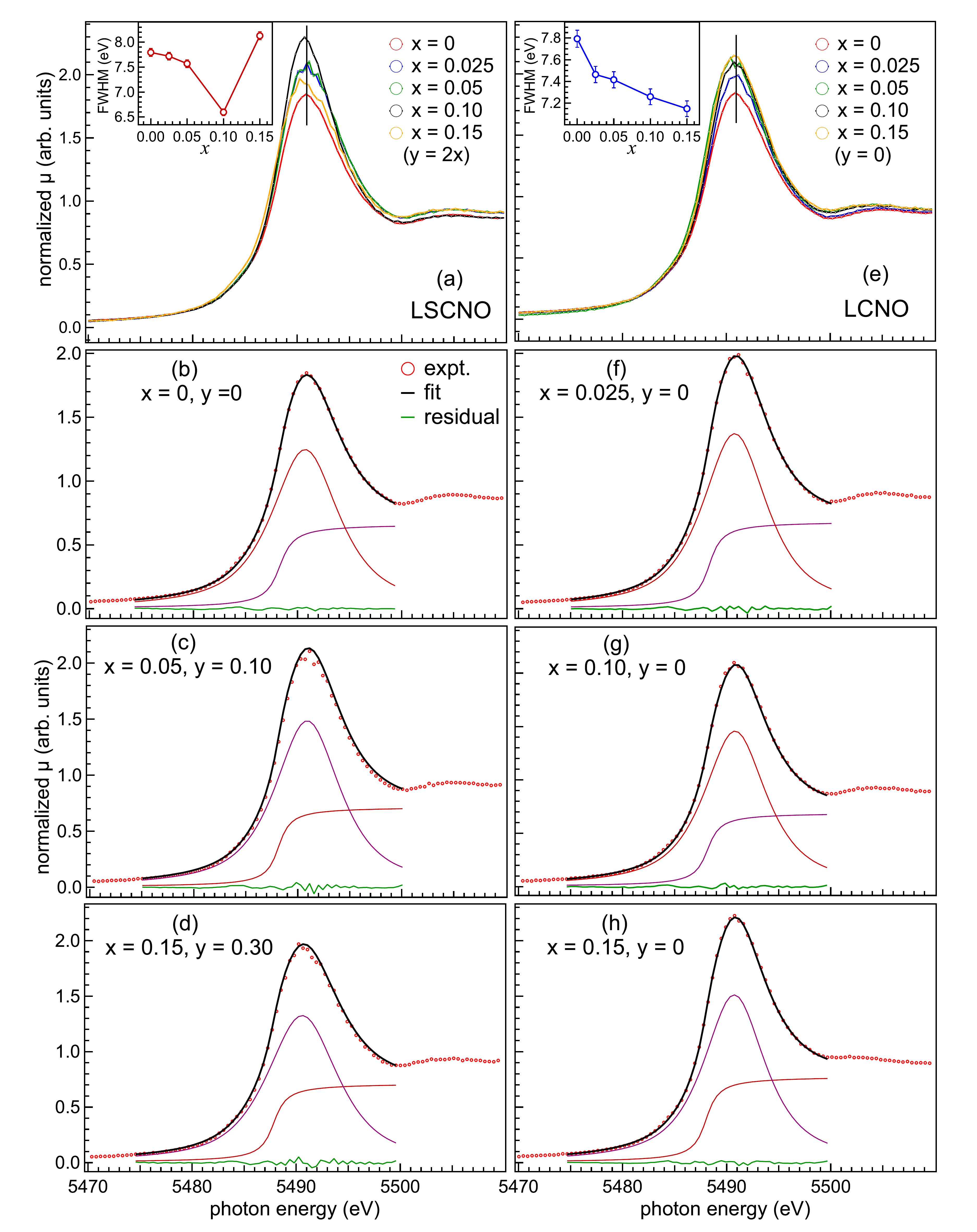}
\renewcommand{\figurename}{Figure}
\caption{The normalized La L$_3$-edge XANES spectra of the La$_{1-y}$Sr$_{y}$Co$_{1-x}$Nb$_x$O$_3$ samples (a) LSCNO, and (e) LCNO ($x=$ 0--0.15); insets in (a, e) panels show the change in the FWHM of the white line with $x$; fitted XANES spectra with one main peak and a step function for the selected (b-d) LSCNO and (f-h) LCNO samples.}
\label{fig:XANES_La-L3}
\end{figure} 

For the detailed investigation, we have deconvoluted and fitted the XANES spectra considering, a term owing to the transitions from Co 1$s$ to the continuum, i.e., step function, and the peak features due to the minor and main resonance transitions \cite{HaasJSSC04}.  In the assumption that the line shape of the XANES continuum step can be determined by the core-hole lifetime, an integration of the pseudo-Voigt function can be utilized as a step function \cite{KoningsbergerJWS88}. The room temperature XANES spectra are fitted in the energy range of 7700--7731~eV  [see Figs.~\ref{fig:XANES_Co_K}(b--d, f--h)], with one step function (modeled as the error function) and four peak features, A, B, C, and D, responsible for the energy transitions explained previously in reference to Figs.~\ref{fig:XANES_Co_K}(a, e). The analysis exhibit that the Gaussian is the best fitted function to model the transitions in the present samples, as followed by Haas {\it et al.} \cite{HaasJSSC04}. The pre-edge features (A and B) are fitted for the LSCNO samples in Figs.~\ref{fig:XANES_Co_K}(b-d), where the relative area and separation remains almost constant and confirm the presence of analogous spin-states in these samples. Since a decrease in the area of peak A can be related to the presence of a small contribution from Co$^{2+}$ states and change in the peak position is related to the enhancement in the r$\rm_{Co-O}$ and Co--O--Co bond-angle. In the case of the LCNO samples, Figs.~\ref{fig:XANES_Co_K}(f--h), the pre-edge features alter with the Nb substitution. The first pre-edge (peak A) shifts towards the lower photon energy due to the enhanced insulating nature of substituted samples, which will result in the decrease of bandwidth near the Fermi level and a stronger interaction with the core-hole potential \cite{IgnatovPRB01}. In such a condition, the quadrupole transition should have a smaller shift to the higher energies. This effect is also validated with the increased broadening of peak A in the pre-edge region. A systematic change in these recorded peaks for LCNO samples can be due to the increased concentration of Co$^{2+}$ ions and a subtle enrichment of electrons in the HS state \cite{ShuklaPRB18, ChangPB03}. We observed a shift in the lower photon energy of around 0.2, 0.7~eV in peak A for $x=$ 0.1 and 0.15 sample, respectively, and a shift towards higher photon energy of 0.3~eV in peak B for $x=$ 0.15 sample. The variation in the peak C (7716~eV) and D (~7726~eV) is directly related to the variation in the oxidation state of Co-ions because these features are related to the final state effects as defined earlier, which do not alter for LSCNO, whereas a subtle variation is observed for the LCNO samples \cite{PandeyJPCM06, ChainaniPRB92, KimPB97}. 

We have measured the La L$_3$-edge spectra, which have the advantage of a longer core-hole lifetime as compared to the K-edge \cite{GlatzelPRB05}, where the Figs.~\ref{fig:XANES_La-L3}(a, e) display the normalized absorption spectra (calibrated with the V-foil) for the LSCNO and LCNO samples, respectively. We observe that the L$_3$-edge exhibits a strong peak around 5490.5~eV, which can be assigned to the transition from 2$p_{3/2}$ to 5$d$ unoccupied states \cite{AsakuraCR19}. The peak position agrees with the reported value for the LCO sample \cite{AsakuraCR19} and there is no significant shift for the LSCNO and LCNO samples, as indicated by the solid vertical lines in Figs.~\ref{fig:XANES_La-L3}(a, e). This indicates that the oxidation state of La cations do not alter and preserved in the 3+ ([Xe]5$d^0$6$s^0$6$p^0$) state despite the Sr$^{2+}$ and Nb$^{5+}$ substitution at A and B sites, respectively, in LSCNO/LCNO samples. The LCO has 12 adjacent oxygen atoms near the La$^{3+}$ cation, which exhibits the narrowest white line in comparison to the other lower coordinated compounds \cite{AsakuraIC14}. For the detailed quantitative investigation of the measured La L$_3$-edge spectra, we have fitted the XANES region with one arctangent step function, which means the electric transition from 2$p_{3/2}$ to continuum states and using a pseudo-Voigt peak shape function, which represents the transition to the unoccupied states. The inset of Figs.~\ref{fig:XANES_La-L3}(a, e) shows the FWHM of the white line for the La L$_3$-edge XANES spectra for all the samples. For the LSCNO samples, the FWHM of the white line decreases up to $x \le$ 0.1, which manifests the decrease in local disorder about the La cations; whereas, a higher value for the $x=$ 0.15 sample indicates higher disorder, which is correlated with the off-limit concentration from the bond-valence sum calculation in ref.~\cite{ShuklaJPCC19}. Interestingly, for the LCNO samples, we observe a monotonous decrease in the FWHM values, which reveals that the local disorder around the La site decreases with Nb substitution. There exist a valid correlation between the FWHM of the white line of L$_3$-edge and the pre-edge peak area of the L$_1$-edge (not shown). Here, the correlation indicates that the hybridization of the $p$-type state into $d$-type states simultaneously allows electric dipole transition in the pre-edge region at the L$_1$-edge and broadens the $d$-type final state to result in the broadening of the white line at the L$_3$-edge \cite{AsakuraIC14}. 

\subsection{\noindent ~EXAFS Analysis}

\begin{figure*}
\includegraphics[width=7.3in]{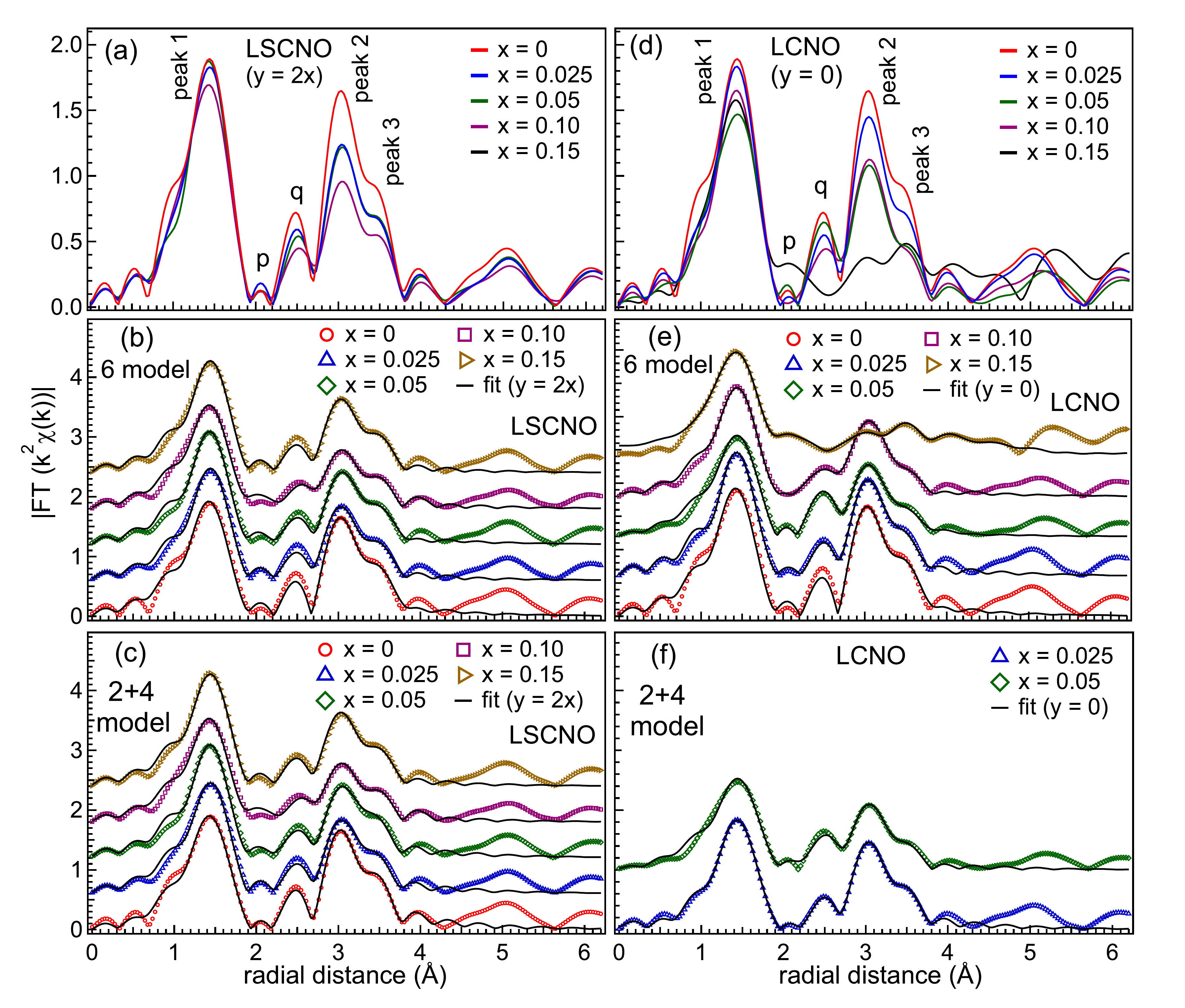}
\renewcommand{\figurename}{Figure}
\caption{A comparison of the Fourier transform (FT) of the EXAFS spectra ($k^2\chi(k)$) measured at Co K-edge as a function of radial distance R for the La$_{1-y}$Sr$_{y}$Co$_{1-x}$Nb$_x$O$_3$ samples (a) LSCNO and (b) LCNO, respectively. The FT of the EXAFS spectra is simulated with the (c) 6-model and (d) 2+4-model for LSCNO samples. The simulated FT of the EXAFS spectra with the (e) 6-model and (b) 2+4 model for the LCNO ($x=$ 0.025 and 0.05) samples, all the spectra are shifted vertically with a constant factor of 0.6 for better presentation.}
\label{fig:EXAFS_Co_K-edge}
\end{figure*}

\begin{table*}
\renewcommand{\tablename}{Table}
		\renewcommand{\thetable}{\arabic{table}}
\caption{The fitting parameters of EXAFS curves for the La$_{1-y}$Sr$_{y}$Co$_{1-x}$Nb$_x$O$_3$ (LSCNO, $y=$ 2$x$) samples obtained using the two different models at the Co K-edge.}
\vskip0.2cm
\begin{tabular}{|c|c|c|c|c|c|c|c|c|c|c|}
\hline
\rule{0pt}{15pt}&\multicolumn{2}{|c|}{$x=$ 0}&\multicolumn{2}{|c|}{$x=$ 0.025}&\multicolumn{2}{|c|}{$x=$ 0.05}&\multicolumn{2}{|c|}{$x=$ 0.1}&\multicolumn{2}{|c|}{$x=$ 0.15}\\
\hline
\rule{0pt}{20pt}&Co--O & DW factor &Co--O & DW factor &Co--O & DW factor &Co--O & DW factor &Co--O & DW factor \\
\rule{0pt}{15pt}&(\AA)&(x10$^{-3}$\AA$^2$)&(\AA)&(x10$^{-3}$\AA$^2$)&(\AA)&(x10$^{-3}$\AA$^2$)&(\AA)&(x10$^{-3}$\AA$^2$)&(\AA)&(x10$^{-3}$\AA$^2$)\\
\hline
\rule{0pt}{15pt}6 model&6$\times$1.910(2)&3.9(5)&6$\times$1.912(4)&3.5(3)&6$\times$1.916(4)&2.9(5)&6$\times$1.913(5)&5.0(6)&6$\times$1.921(2)&2.5(4)\\
\hline
\rule{0pt}{15pt}R factor&\multicolumn{2}{|c|}{0.011}&\multicolumn{2}{|c|}{0.010}&\multicolumn{2}{|c|}{0.011}&\multicolumn{2}{|c|}{0.009}&\multicolumn{2}{|c|}{0.006}\\
\hline
\rule{0pt}{15pt}2+4 model&2$\times$1.861(3)&3.2(4)&2$\times$1.856(5)&0.7(3)&2$\times$1.862(4)&0.9(3)&2$\times$1.823(4)&0.1(1)&2$\times$1.827(1)&2.4(3)\\
\hline
\rule{0pt}{15pt}&4$\times$1.921(3)&4.5(5)&4$\times$1.928(2)&0.1(1)&4$\times$1.931(5)&2.3(3)&4$\times$1.947(2)&1.0(2)&4$\times$1.949(3)&0.1(1)\\
\hline
\rule{0pt}{15pt}average&1.901&-&1.904&-&1.908&-&1.906&-&1.908&-\\
\hline
\rule{0pt}{15pt}R factor&\multicolumn{2}{|c|}{0.003}&\multicolumn{2}{|c|}{0.006}&\multicolumn{2}{|c|}{0.011}&\multicolumn{2}{|c|}{0.008}&\multicolumn{2}{|c|}{0.002}\\
\hline
\end{tabular}
\end{table*}

Now we examine the local structure in detail by performing the quantitative analysis of the Fourier transformation (FT) of the  EXAFS spectra using the Artemis program developed by Ravel {\it et al.} under the Demeter package \cite{RavelJSR05}. In Figs.~\ref{fig:EXAFS_Co_K-edge}(a, d), we present the magnitude of Fourier transform (FT) of the EXAFS spectra ($k^2\chi(k)$) in the R--space, which are performed in a Hanning window of 3--11.2~\AA$^{-1}$ k--range, a R$_{bkg}$ background of 1, and 0.2~\AA$^{-1}$ Gaussian broadening \cite{RavelJSR05} for the series of both LSCNO and LCNO samples. Interestingly, in the case of the LCO (red curve) sample we can see three main peaks in the FT of the EXAFS spectra centered near the R-values of 1.5 (peak 1), 3.0 (peak 2), and 3.5~\AA~(peak 3), which can be attributed to the Co-O, Co-La, and Co-Co bonds, respectively. However, the small peaks marked as $p$(2.0~\AA) and $q$(2.5~\AA) are because of the weaker contributions from the Co-O and Co-La bonding, which are also termed as spectral leakage \cite{PandeyJPCM06}. The main peak near 1.5~\AA~(peak 1) corresponds to the octahedral coordination with the nearest neighbor oxygen atoms. The octahedral distortion caused by the Co-O bonds reflects in the first shell of FT of the EXAFS spectra \cite{HaasJSSC04, PandeyJPCM06}. For the LSCNO samples, there is no significant change in the shape of the EXAFS spectra with cationic substitution, while the peaks that appeared for the LCO ($x=$ 0) sample alter their intensity such that the peaks 1, 2, and 3 decrease in intensity up to $x=$ 0.1, and then increase for the $x=$ 0.15 sample. This observation manifests that the neighboring structure of Co-La  and Co-Co/Nb bonds is modified with the co-substitution (Sr and Nb); whereas the Co-O bond is almost unaffected. In the case of the LCNO samples, the rhombohedral structure is preserved up to $x=$ 0.05, where we can observe a decrease in the intensity for higher Nb concentration without affecting the peak shape, while for $x >$0.05, a structural phase transformation prevails in the system. Here, except for the first peak due to the CoO$_6$ octahedra, we observe that the peaks for the Co--La and Co--Co change drastically with the Nb substitution, which manifests a cationic substitution induced variation in the local structure. We have simulated the FT of the EXAFS spectra in the R-range of 1--3.8~\AA~[see Figs.~\ref{fig:EXAFS_Co_K-edge}(b,c,e,f)] with single scattering paths only to avoid the higher coordination shells. The simulated FT of the EXAFS spectra are not corrected with the back-scattered and central phase shifts. Therefore, the position of the peaks in FT of the EXAFS curves do not reflect the real distances of neighboring atoms as there is a phase shift of [2$\delta_c(k)$+$\delta_i(k)$] between the real values and the spectra shown here. 

For the curve fitting of the FT of the EXAFS spectra, we have generated all possible scattering paths up to 5~\AA~radius using the FEFF program in Artemis \cite{RehrRMP00, RavelJSR05} on account of unit cell parameters and atomic positions obtained from the earlier x-ray diffraction analysis using Rietveld refinement \cite{ShuklaPRB18, ShuklaJPCC19}. These generated scattering paths are specified with the degeneracy (coordination number), the interatomic distance, and relative weighted contribution in the scattering amplitude. The first shell between R-values of 1--1.9~\AA~is dependent on the CoO$_6$ octahedra and can be reasonably well fitted using the structural results obtained using FEFF [see Figs.~\ref{fig:EXAFS_Co_K-edge}(b,c,e,f)]. The fitting of the higher shells (R-values of 2.1--3.8~\AA) is carried out to get the information about the next nearest atomic scatterers using the Co--(La/Sr) and Co--(Co/Nb) scattering paths. The width of Co--O pair-distribution function for the parametrization of disorder is determined by fixing the oxygen coordination around the Co atom to 6, and to minimize the free parameters, the first shell fitting is done in the Fourier filtered k-space (1--1.9~\AA~range) considering all equal Co--O bond lengths. The reduction factor S$_o^2$ is observed between 0.8$\pm$0.1 for all the samples. Moreover, the value of $\Delta$E$_o$ relative to the threshold energy governs the value of R and integrated intensity for the individual atomic scattering path. So, the variation in the value of $\Delta$E$\rm_o$ will lead to the ambiguous results from the curve fitting, therefore we have used an analogous value for all the atomic scattering paths, as reported in ref.~\cite{HaasJSSC04}.

\begin{figure*}
\centering
\includegraphics[width=7.3in]{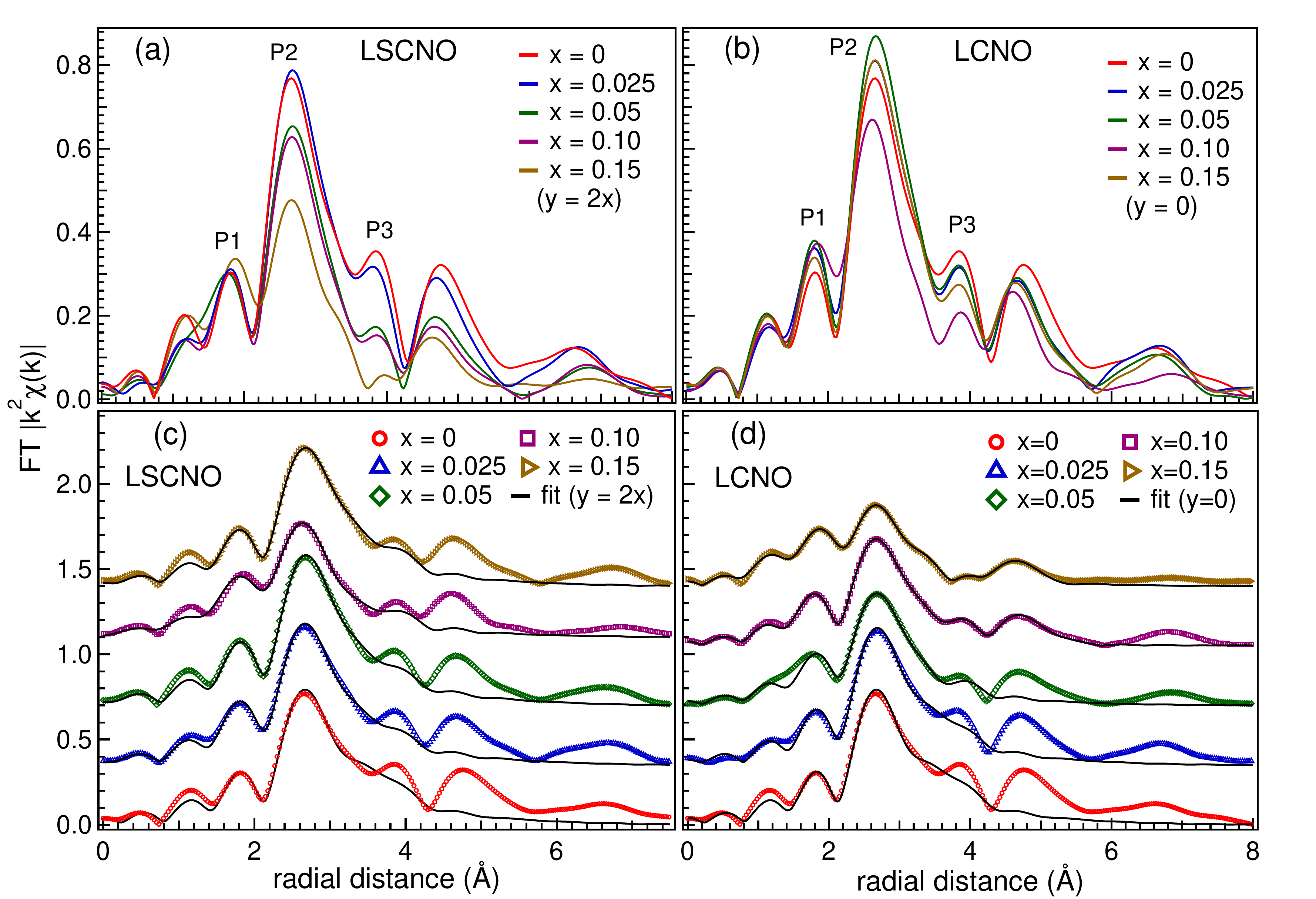}
\renewcommand{\figurename}{Figure}
\caption{A comparison of the Fourier transform (FT) of the EXAFS spectra ($k^2\chi(k)$) of experimentally recorded data at the La L$_3$-edge as a function of radial distance R for the La$_{1-y}$Sr$_{y}$Co$_{1-x}$Nb$_x$O$_3$ samples (a) LSCNO and (b) LCNO. (c, d) The experimental and simulated FT of $k^2\chi(k)$ for the LSCNO and LCNO samples, respectively, each spectrum is shifted vertically by a constant factor of 0.35 for better presentation.}
\label{fig:EXAFS_La-L3}
\end{figure*}

Note that the average structure inferred from the x-ray diffraction analysis of the LCO sample is rhombohedral, implying six equal Co--O bonds of the CoO$_6$ octahedron. In the FT of the EXAFS spectra, we have fitted the first shell using the scattering from the nearest O atoms consist of six equal Co--O bond lengths, which is coined as the 6-model [see Figs.~\ref{fig:EXAFS_Co_K-edge}(b, e)]. However, in the curve fitting performed with the 6 model, we found that the quality of fitting estimated by the R-factor in Artemis is not good [see Table-1]. In order to find the different local distortions in the CoO$_6$ octahedron, various models/hypotheses can be utilized; for example, the site-disorder for atoms, JT distortion using (2+4 or 4+2 model), and variation in the coordination number of nearest neighbors. In this context, the IS state induced Jahn-Teller (JT) distortion is reported in the LCO using different experimental tools like phonon modes \cite{YamaguchiPRB96}, thermal conductivity \cite{YanPRB03}, neutron pair distribution \cite{SundaramPRL09}, and x-ray absorption \cite{SundaramPRL09, PandeyJPCM06, HaasJSSC04}. This JT distortion in LCO at room temperature appears due to the partial occupation of degenerate e$_g$ states in the IS state \cite{KorotinPRB96, HaasJSSC04}. Louca {\it et al.} presented that the distorted CoO$_6$ octahedra at the local level, in the LCO sample, possess four short and two long bond lengths at 100~K in the IS state \cite{LoucaPRB97}. Here we follow a similar idea of lattice distortion and performed the curve fitting using the 2+4 model, where Co--O bond-lengths are divided into two sets of bond lengths in a group of two short bonds and four long bonds, respectively. In the curve fitting using the 2+4 model we first utilized the R range of 1-1.9~\AA, which involves the first peak in the EXAFS spectra [see Fig.~\ref{fig:EXAFS_Co_K-edge}]. For fitting of the higher coordination shell within analogous settings, we fix the Co--O bond lengths in both shells and then get the values of fitted parameters without altering the values of Co--O bonds obtained from the lower coordination shells. The FT of the EXAFS spectra of the LSCNO sample are well fitted with the 2+4 model, and the results specify that there is an improvement in fitting for the higher k-region, which in turn reflects in the FT of the EXAFS spectra, as shown in Figs.~\ref{fig:EXAFS_Co_K-edge}(c) and can be estimated by the R-factor values in Table-1. The output parameters obtained from the analysis of the EXAFS spectra using two models are compared in Table-1 for the LSCNO samples. There is an additional possibility to look for the distortion using the 4+2 model too, which involves the similar practice as for the 2+4 model, but two groups of four short and two long bonds. However, the curve fitting using the 4+2 model does not ameliorate the fitting of data (not shown), hence not considered. Furthermore, as all the LSCNO samples exhibit the rhombohedral structure, we follow the 2+4 model in the EXAFS curve fitting analogous to the $x=$ 0 sample and present in Fig.~\ref{fig:EXAFS_Co_K-edge}(c). Interestingly, we observe that the JT distortion present in the $x=$ 0 sample is very stable despite Sr and Nb substitution in the LSCNO samples. This outcome further validates the presence of Co$^{3+}$ ions in the IS state at room temperature for all the LSCNO samples. In this case, owing to the populated e$_g$ state, the possibility of the Jahn-Teller distortion is higher because of the degenerate ground state \cite{KorotinPRB96}. However, as suggested by Louca {\it et al.} the occupancy of the d$_{x^2-y^2}$ in the IS state changes the Co--O environment and will be more pronounced as compared to the occupancy of d$_{z^2}$ orbital \cite{LoucaPRB97}. Therefore, the different occupancies of the $d-$orbitals in the e$_g$-state are a consequence of the JT-distortion in these samples. Moreover, the fitting results exhibit that  the local structure probed by the EXAFS analysis is different from the overall crystal structure seen by the XRD measurements.

\begin{table*}
\renewcommand{\tablename}{Table}
		\renewcommand{\thetable}{\arabic{table}}
\caption{The EXAFS curve fitting parameters, bond-length (\AA), DW factor ($\times$10$^{-3}$\AA$^2$) for the La$_{1-y}$Sr$_{y}$Co$_{1-x}$Nb$_x$O$_3$ samples (LSCNO; $y=$ 2$x$ and LCNO; $y=$ 0) measured at the La L$_3$-edge.}
\vskip0.2cm
\setlength{\tabcolsep}{7pt}
\begin{tabular}{|c|c|c|c|c|c|c|c|}
\hline
\rule{0pt}{15pt}&{LCO}&{LSCNO}&{LSCNO}&{LSCNO}&{LSCNO}&{LCNO}&{LCNO}\\
\rule{0pt}{15pt}&{$x=$ 0}&{$x=$ 0.025}&{$x=$ 0.05}&{$x=$ 0.1}&{$x=$ 0.15}&{$x=$ 0.025}&{$x=$ 0.05}\\
\hline
\rule{0pt}{15pt}3$\times$La--O1&2.443(5)&2.451(4)&2.454(2)&2.460(3)&2.465(5)&2.452(4)&2.457(3)\\
\rule{0pt}{15pt}DW factor&11.5(7)&5.5(4)&6.6(4)&8.0(5)&5.8(7)&9.8(6)&7.9(5)\\
\rule{0pt}{15pt}6$\times$La--O2&2.672(4)&2.683(3)&2.686(5)&2.702(4)&2.711(8)&2.684(6)&2.687(4)\\
\rule{0pt}{15pt}DW factor&11.5(7)&5.5(4)&6.6(4)&8.0(5)&5.8(7)&9.8(6)&7.9(5)\\
\rule{0pt}{15pt}3$\times$La--O3&2.947(5)&2.936(3)&2.934(5)&2.922(7)&2.908(5)&2.919(4)&2.913(4)\\
\rule{0pt}{15pt}DW factor&8.6(5)&6.1(3)&7.5(4)&1.5(3)&5.1(7)&6.8(5)&1.3(2)\\
\hline
\rule{0pt}{15pt}6$\times$La--Co&3.292(4)&3.307(6)&3.312(3)&3.323(5)&3.382(6)&3.315(4)&3.314(6)\\
\rule{0pt}{15pt}DW factor&5.3(6)&2.5(4)&1.8(3)&4.9(5)&5.6(7)&2.9(3)&2.9(2)\\
\hline
\rule{0pt}{15pt}6$\times$La--La&3.758(3)&3.763(6)&3.763(5)&3.765(4)&3.788(6)&3.743(2)&3.782(7)\\
\rule{0pt}{15pt}DW factor&10.8(7)&12.6(5)&10.6(5)&11.5(4)&8.3(8)&2.1(3)&8.2(6)\\
\hline
\rule{0pt}{15pt}R factor&0.018&0.017&0.015&0.019&0.011&0.015&0.014\\
\hline
\end{tabular}
\end{table*}

In the case of the LCNO samples, from the previous XRD analysis, we know that for $x\le$ 0.05 samples structure can be modeled using a rhombohedral space group (R-3c), and a fixed proportion of the rhombohedral (R-3c) and orthorhombic (Pbnm) space groups were used for the $x >$ 0.05 samples \cite{ShuklaPRB18}. In order to generate the atomic scattering paths in the FEFF, we have considered the initial phase fraction from the Rietveld refinement of the XRD patterns \cite{ShuklaPRB18} and utilized them to model the EXAFS curve fitting. The fitted FT of the EXAFS spectra for the LCNO samples ($x=$ 0.025--0.15) are presented in the Figs.~\ref{fig:EXAFS_Co_K-edge}(e, f). In our previous report on the LCNO samples, the values of effective magnetic moment advocate that Co$^{3+}$ ions exist in a mixture of IS (10\%) and HS (90\%) state for the $x=$ 0.025 sample \cite{ShuklaPRB18}. Here, we perform the FT of the EXAFS curves using both the 6-model and 2+4 model and found the values of R-factor 0.014 and 0.006, respectively. The improved fitting using the 2+4 model [see Fig.~\ref{fig:EXAFS_Co_K-edge}(f)] strongly suggest that the Co$^{3+}$ ions present  in the IS state are JT-active due to the partial occupation of e$_g$ state \cite{KorotinPRB96, HaasJSSC04, PandeyJPCM06}. Therefore, it is an important finding as the orbital magnetic moment may not be quenched in the $x=$ 0.025 sample due to the presence of the Co$^{2+}$ state, which also reflects in the higher value of the effective magnetic moment ($\mu_{eff}$) \cite{ShuklaPRB18}. Further in this line, to investigate the possibility of JT-distortion in the $x=$0.05 sample, we have also fitted the FT of the EXAFS spectra using the 2+4 model and presented in Fig.~\ref{fig:EXAFS_Co_K-edge}(f), where we found that the quality of curve fitting for this sample is not improved as compared to the 6-model. Hence, a reasonable good fit of the $x=$ 0.05 sample using the 6-model exhibits that the presence of a Co$^{3+}$ in the HS state and a higher concentration of Co$^{2+}$ ions are responsible for the absence of JT-effect \cite{ShuklaPRB18}. Similarly, the FT of the EXAFS spectra for the $x=$ 0.1 and 0.15 sample is fitted with 6-model considering a combination of rhombohedral (74\%) \& orthorhombic (26\%), and orthorhombic (58\%) \& monoclinic (42\%) phases, for the $x=$ 0.1 and 0.15 samples, respectively. These fitted FT of the EXAFS spectra are presented in Fig.~\ref{fig:EXAFS_Co_K-edge}(e).

In order to understand the behavior around the La L$_3$-edge, we perform the curve fitting of the EXAFS spectra using the FT in the R-space from the k-range of 3--8.5~\AA$^{-1}$. In the EXAFS curve fitting, the maximum number of independent parameters (N$\rm_I$) are limited by the formula $\frac{2\Delta k\Delta R}{\pi}$+2, where $\Delta$k and $\Delta$R define the range of electron momentum and fitted radial interval, respectively \cite{SternPRB93}. In Figs.~\ref{fig:EXAFS_La-L3}(a, b) we show a comparison of the FT of the EXAFS spectra for the LSCNO and LCNO samples , respectively. In these samples, the first peak near 1.8~\AA~(P1) corresponds to the nearest neighbor O atoms around La absorbers. The La atoms are 12 fold coordinated among the three groups of La--O bond-lengths. The main peak (P2) near 2.7~\AA~is attributed to the contribution from the nearest Co scatterers, i.e., the smallest La--Co bond-length. The peak near 3.7~\AA~(P3) is attributed to the next nearest La atoms, which is related to the La--La bond-lengths. The change in the FT of EXAFS spectral intensity is related to the change in the distance (bond-length) and bond-angle of scatterers from the absorber, therefore a change in these directly reflect the variation of the EXAFS spectral intensity. For the LSCNO samples, the FT of the EXAFS spectra of $x=$ 0 and 0.025 are similar; however, the magnitude decreases for higher values of the $x$ and the lowest value is found for the $x=$ 0.15 sample [see Fig.~\ref{fig:EXAFS_La-L3}(a)]. Moreover, in the case of the LCNO samples, a comparison of FT of the EXAFS spectra shows an increment in the intensity of the most intense P2 peak with Nb substitution except $x=$ 0.1 sample, as shown in Fig.~\ref{fig:EXAFS_La-L3}(b). Note that, in contrast to the Co K-edge spectra of the LCNO samples for $x >$ 0.05, there is no significant change in the FT of the EXAFS spectra, which suggest that the Nb substitution does not alter the local structure around the La site. The FT of $k^2\chi(k)$ is fitted in the R-range of 1.3--4.1~\AA~considering the unit cell structural parameters obtained from the Rietveld refinement of the diffraction patterns in refs.~\cite{ShuklaJPCC19, ShuklaPRB18}. 
\begin{figure}
\centering
\includegraphics[width=3.55in]{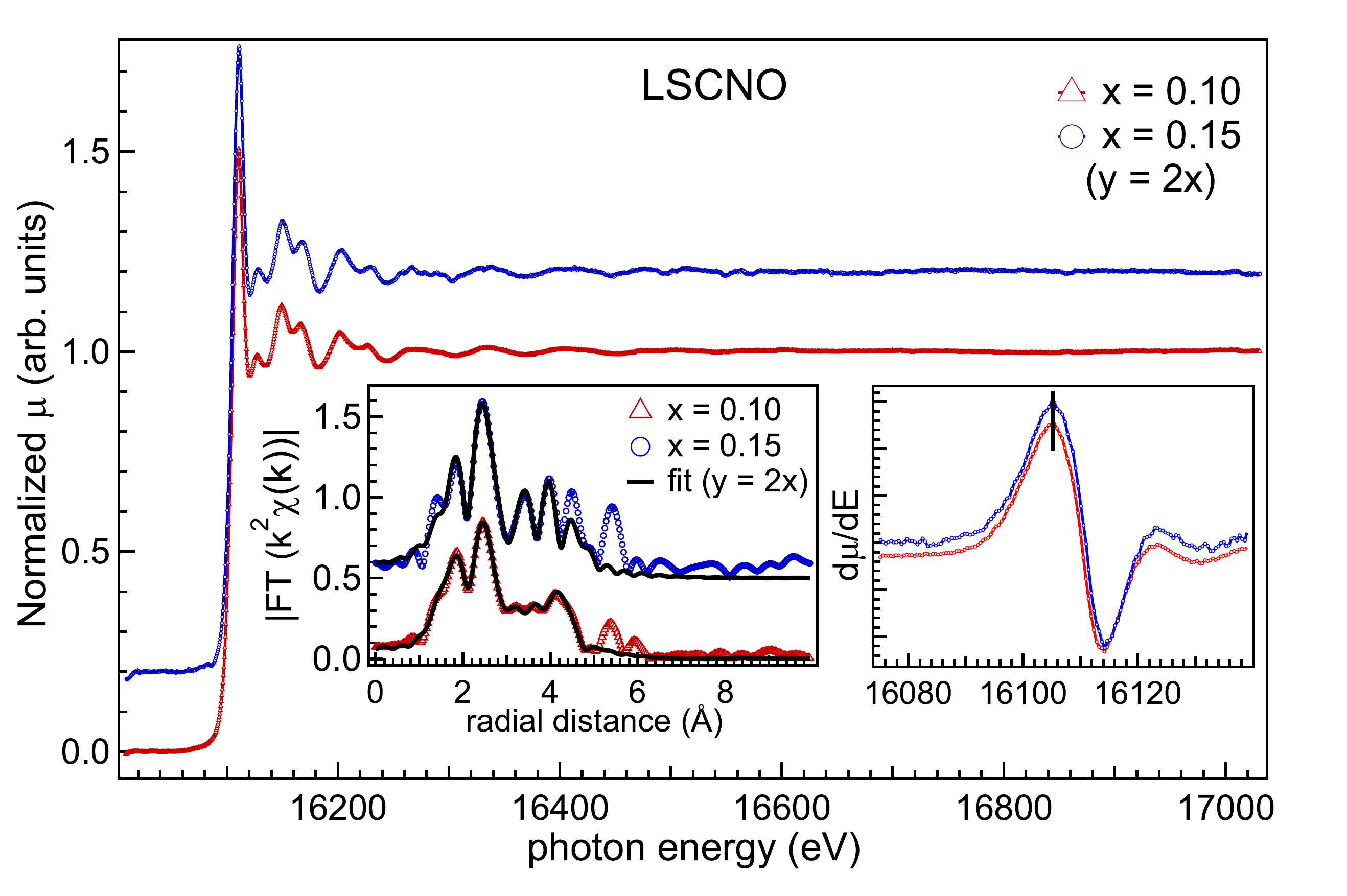}
\renewcommand{\figurename}{Figure}
\caption{The normalized experimental Sr K-edge absorption spectra of the La$_{1-y}$Sr$_{y}$Co$_{1-x}$Nb$_x$O$_3$ samples ($x=$ 0.1 and 0.15; $y=$ 2$x$) samples, vertically shifted by a constant factor of 0.2 for clear presentation. The left and right insets present the simulated and experimental Fourier transform of the EXAFS spectra and first-order differential near inflection around the edge jump.}
\label{fig:Sr_K-edge}
\end{figure}
For a detailed insight into local coordination at La-site, we perform the curve fitting of FT of the EXAFS spectra for the LSCNO and LCNO samples and the output parameters are presented in Table-2. For the LSCNO samples, we found that the long-bonds (La--O) shrink and the other two groups of bonds elongate with cationic substitution ($x$), which is in agreement with neutron diffraction results \cite{ShuklaJPCC19}. A decrease in the La--O bond length with the hydrostatic pressure to preserve the crystal structure is reported in ref.~\cite{CaponePSSA19}, and in the present case, we observe a similar behavior despite the large size Sr and Nb substitution in the lattice. Moreover, the fitting of the second shell corresponding to the La--Co bonds exhibits a monotonous enhancement in the La--Co bond-lengths with the cationic substitution. Here, we have fitted the FT of the EXAFS spectra only for the nearest La scatters and we observed an increase in the six-fold degenerate La--La bond length with the $x$. In the case of the LCNO samples, the behavior is similar to the LSCNO samples up to $x=$ 0.05 owing to the similar rhombohedral structure. However, the La--Co bond lengths follow a different trend for the samples with higher Nb concentration. 

Finally, we present the normalized Sr K-edge absorption spectra in Fig. \ref{fig:Sr_K-edge} for the LSCNO sample having higher Sr concentration, i.e., the $x=$ 0.10 and 0.15 samples. The spectra for $x<$ 0.10 had a smaller signal to noise ratio due to the lower concentrations of Sr, hence not presented here. A comparison of the first-order differential at Sr K-edge with reported data confirm that Sr ions exist in the 2+ oxidation state for both samples \cite{BazinJSR14}, as highlighted with a solid black-line in the right inset of Fig.~\ref{fig:Sr_K-edge}. On the other hand, an enhancement in the FWHM of the main peak in XANES at a higher concentration is observed, which manifests a change in the local environment due to the cationic substitution of Sr and Nb despite no change of valence state at the A-site. The analysis of the FT of the EXAFS spectra in the R-space is shown in the left inset of Fig.~\ref{fig:Sr_K-edge} for the $x=$ 0.1 and 0.15 samples where the peaks $\approx$2.5~\AA~are associated with the nearest O scatterers, and peaks between 3-4~\AA~appear due to the scattering from the next nearest Co and La atomic scatterers. The EXAFS curve fitting at the Sr K-edge exhibits that there are three groups of Sr-O bond lengths and six-fold degenerate Sr--Co and Sr--Sr bonds analogous to the results obtained from the analysis of La L$_3$-edge. The spectra are shifted vertically with a constant factor for a better visual presentation. In addition to this, the Nb K-edge spectra were recorded at room temperature for the LCNO ($x=$ 0.1 and 0.15) samples. In order to check the oxidation state of Nb ions we compare the XANES spectra with the reference spectrum of Nb$_2$O$_5$ after calibration using the Nb foil (not shown here). We found that the maxima in the first-order derivative of spectra match with the energy position for the reference spectrum, which confirms the 5+ oxidation state of Nb ions in the samples.

\section*{\noindent ~Conclusion}

We have investigated the local structure and electronic properties of Sr and Nb substituted LaCoO$_3$ (LCO) using x-ray absorption measurements at room temperature. The XANES spectra at Co K-edge for the LSCNO samples confirm the 3+ oxidation state of Co ions; however, the Nb substitution in LCO leads to the reduction in the average valence of the Co ions due to the emergence of Co$^{2+}$ ions. Interestingly, the EXAFS curve fitting of the LSCNO samples at the Co K-edge manifests that the Co$^{3+}$ ions are Jahn-Teller (JT) active and result in the two sets of Co--O bond-lengths, which indicates the presence of IS state. However, for the LCNO, only the $x=$ 0.025 sample was found to be JT active, whereas the $x \ge$ 0.05 samples possess the local symmetry in agreement with the XRD results. The XANES analysis of La L$_3$-edge spectra confirms the trivalent state of La ions and a monotonous decrease of FWHM for the white line reveals a decrease in the local disorder with Sr and Nb substitution. The simulated FT of the EXAFS spectra at La L$_3$-edge exhibit three sets of La--O bond lengths (long, small, and intermediate), which show a systematic change with increasing the substitution. Further, the absorption measurements of the Sr and Nb K-edges confirm their oxidation state to be in 2+ and 5+, respectively.

\section*{\noindent ~Acknowledgment}

RS and AK gratefully acknowledge the DST-Inspire and UGC, India for fellowship, respectively. RS thanks Dr. Ashok K. Yadav for useful discussion on XAS analysis. We thank SERB-DST for financial support through Early Career Research (ECR) Award (project reference no. ECR/2015/000159). We also acknowledge the CRS project no. CSR-IC-ISUM-36/CRS-319/2019-20/1371 for collaborative research work.

\end{document}